# Electric-Field Ionization of Gallium Acceptors in Germanium Induced by Single-cycle Terahertz Pulses

Y. Mukai,[1] H. Hirori,[2,*] and K. Tanaka[1,2,†]

[1] *Department of Physics, Graduate School of Science, Kyoto University, Sakyo-ku, Kyoto 606-8502, Japan*

[2] *Institute for Integrated Cell-Material Sciences (WPI-iCeMS), Kyoto University, and JST-CREST, Kyoto 606-8501, Japan*

## Abstract

The electric field ionization of gallium acceptors in germanium was studied by using terahertz time-domain spectroscopy after single-cycle terahertz pulse excitation. As the peak electric field of the excitation pulse increases, the distinct absorptions due to acceptor transitions centered at 2.0 and 2.2 THz decrease, and simultaneously, absorption emerges in the lower frequency region. These behaviors clearly show that the terahertz pulse ionizes neutral acceptors. The electric field dependence of the released hole density is well reproduced by a model assuming direct field-assisted tunneling of acceptors.

* hirori@icems.kyoto-u.ac.jp

† kochan@scphys.kyoto-u.ac.jp

The interactions of impurity states in semiconductors with terahertz (THz) electromagnetic waves have attracted considerable attention because of their importance in fundamental physics and technological applications. Shallow impurities have a set of discrete electronic bound states similar to the Rydberg series of the free hydrogen atom. The specific excitation energies between the states lie in the few meV or THz spectral region.[1] The ability to induce nonlinear interactions of impurity states with high-power THz pulses has opened up the possibility of coherent manipulation of quantum bits in ubiquitous semiconductors.[2,3]

For ultrafast coherent manipulation of impurity states, the duration of the controlling THz pulse should be shorter than the coherence time.[4] Simultaneously, in order to keep the pulse area large enough for inducing a population inversion, the peak electric field of the controlling pulse must be intense. In the high electric field regime, the nonlinear process of ionizing impurities under THz pulse irradiation is crucial for coherent manipulation. So far, ionization phenomena such as multi-photon ionization in shallow impurity states and phonon-assisted tunneling in deep impurity states have been studied in the weak electric field regime by means of nanosecond far-infrared laser pulse irradiation.[5,6] To be able to study the high electric field regime, the nonlinear process of ionizing impurities under THz pulse irradiation is crucial for coherent manipulation. The recent development of ultra-intense THz laser systems generating phase-stable transients has enabled us to study coherent THz manipulation[7-9] and fascinating THz nonlinear phenomena in various materials.[10-12] It is claimed that the field ionization process may play a key role in these nonlinear phenomena. However, the ionization process under instantaneous high electric fields is not well understood.

In this paper, we report on the nonlinear field ionization process of gallium acceptors in germanium (Ge:Ga) studied by using single-cycle THz pulse excitation and successive THz spectroscopy. A first THz-pump pulse induced a change in the complex dielectric constant of the sample, which was measured by THz time-domain spectroscopy using a second THz-probe pulse with a fixed time delay. As the THz-pump electric field increases, the absorption peaks corresponding to the internal acceptor transitions centered at 2.0 and 2.2 THz disappear and a free carrier response appears in the lower frequency region. An analysis of these data based on the Drude-Lorenz model revealed that the THz-pump pulse ionizes the neutral acceptors and releases holes from the acceptors. The pump electric field dependence of the released free hole density is in good agreement with a theoretical calculation assuming a direct field-assisted tunneling process.

Figure 1 shows the experimental setup of the THz excitation and spectroscopy. THz pulses were generated by optical rectification of femtosecond laser pulses in a $LiNbO_3$ crystal by using the tilted-pump-pulse-front scheme.[13-15] An amplified Ti:sapphire laser (repetition rate 1 kHz, central wavelength 780 nm, pulse duration 100 fs, and 4 mJ/pulse) was used for the light source. As shown in the inset of Fig. 1, the generated THz pulse was split into pump and probe pulses on the same axis by means of internal reflection between a pair of wire-grid polarizers (WG1 and WG2). The ratio between the pump and probe electric fields is determined by the relative angle $\theta$ between the polarization axes of WG1 and WG2 and is given by $1:\sin^2\theta$, where the polarization axis of WG2 is parallel to THz electric field. The delay between two pulses of 200 ps corresponds to the round-trip time between the two wire-grid polarizers (WG1 and WG2), allowing us to neglect the pump electric field effect on the spectra observed by the probe pulse. The other pair of wire-grid polarizers (WG3 and WG4) was used to change the field amplitudes of the pump and probe pulses without modifying their waveforms. In addition, in all the experiments, the electric fields of the probe pulses were kept sufficiently below 1 kV/cm, enabling us to rule out spectral modulation due to the probe pulse incidence.[16-18] The sample of gallium-doped germanium (Ge:Ga) crystal (thickness 500 μm) was put in a liquid-He cryostat. The polarizations of the pump and probe pulses were the same and along the <110> direction of the sample.

The electro-optic (EO) sampling technique with a 1-mm-thick ZnTe crystal was used to detect the electric field of the pump and probe pulses transmitted through the sample. Figures 2(a) and 2(b) show the measured temporal profile of the THz field and its power spectrum, and Figure 2(c) shows the real part of the conductivity $\sigma_1$ obtained by THz time-domain spectroscopy without a pump excitation.[19] The spectral analysis was conducted in the 0.3-2.4 THz range. The absorption peaks at 2.0 and 2.2 THz are internal acceptor transitions of $1\mathrm{S}_{3/2}(\Gamma_8^+) \rightarrow 2\mathrm{P}_{5/2}(\Gamma_8^-)$ and $1\mathrm{S}_{3/2}(\Gamma_8^+) \rightarrow 2\mathrm{P}_{5/2}(\Gamma_7^-)$, respectively.[20,21]

Figures 3 show the real part of the dielectric constant $\varepsilon_1$ and the conductivity $\sigma_1$ after THz-pump pulse excitations. The maximum peak field of the pump pulse used in this experiment was estimated to be 13.5 kV/cm inside the sample by calibrating the EO sampling signal considering Fresnel loss.[22] As shown in the figures, as the pump electric field increases, the conductivity around 2 THz, i.e., the acceptor absorption, decreases and eventually vanishes. Below 1 THz, the conductivity increases, and at the same time, the real part of the dielectric constant decreases, implying the emergence of a free carrier response.

The phenomenological Drude-Lorentz model is able to reproduce the obtained spectra. According to this model, the complex dielectric constant $\tilde{\varepsilon}(\omega) = \varepsilon_1(\omega) + i\sigma_1(\omega)/\varepsilon_0\omega$ is given by:

$$\tilde{\varepsilon}(\omega) = \varepsilon_{\mathrm{b}} + \frac{\omega_{\mathrm{p}}^2}{-\omega^2 - i\omega\gamma} + \frac{e^2 N_{\mathrm{b}} \varGamma_1}{m_0 \varepsilon_0} \sum_j \frac{f_j}{\omega_j^2 - \omega^2 - i\omega\gamma_j} \, , \qquad (1)$$

where $\omega$ is angular frequency, $e$ is the elementary charge, $m_0$ is the mass of the free electron, and $\varepsilon_{\mathrm{b}}$=15.3 is the background dielectric constant of germanium.[23] The second term of the Drude dispersion denotes the free carrier response and is characterized by the carrier scattering rate $\gamma$ and the plasma frequency $\omega_{\mathrm{p}} = \sqrt{N_{\mathrm{f}} e^2 / m_{\mathrm{hh}}^* \varepsilon_0}$, with the free hole density $N_{\mathrm{f}}$, effective mass of the heavy hole $m_{\mathrm{hh}}^*$=0.35 $m_0$,[24] and vacuum permittivity $\varepsilon_0$. The third term of the Lorentz dispersion represents the internal transitions between the acceptor levels and is characterized by $\omega_j$, $f_j$ and $\gamma_j$, which are respectively the resonant frequency, oscillator strength, and damping constant of the $j$th ($j$=1, 2) Lorentz oscillator. Notice that $j$ (=1 and 2) corresponds to the two acceptor transitions of $1S_{3/2}(\varGamma_8^+) \rightarrow 2P_{5/2}(\varGamma_8^-)$ and $1S_{3/2}(\varGamma_8^+) \rightarrow 2P_{5/2}(\varGamma_7^-)$ shown in the inset of Fig. 2(c). The oscillator strengths $f_1$ and $f_2$ are given as 0.046 and 0.033, respectively.[23] $N_{\mathrm{b}}$ is the bound hole density, and $\varGamma_1$=13.36 is the Luttinger valence band parameter.[23,25,26]

The dashed lines in Figs. 3 are the theoretical curves obtained by least-squares fitting of the experimental data curves to Eq. (1) using $N_{\mathrm{b}}$, $\omega_j$ and $\gamma_j$ as fitting parameters. The fitted value $N_{\mathrm{b}}(=N_{\mathrm{b}}^0)$ for the no-THz-pump case shown in Figs. 3(a) and 3(b) is deduced to be $\sim$1.4$\times$10$^{15}$ cm$^{-3}$ by using Eq. (1), and is comparable to the acceptor concentration obtained from the electric resistivity measurement. (The uncertainties are typically ±20%, owing mainly to excessive attenuation caused by the sample being too thick for the weak higher-frequency components above 1.5 THz and the error in the multiparameter fitting.) In addition, the fitted value of $N_{\mathrm{f}}$ $\sim$1.3$\times$10$^{15}$ cm$^{-3}$ for the maximum electric field case in Figs. 3(e) and 3(f) is in good agreement with $N_{\mathrm{b}}^0$ obtained for the no-THz-pump case.[27] These results suggest that the incident THz pulses ionize the acceptor impurities, and the free holes released from the acceptors contribute to the free carrier response described by the Drude dispersion.

Figure 4 shows the THz-pump electric field dependence of the estimated bound hole density $N_b$. The bound hole density decreases with increasing electric field. The field dependence shows that the acceptor ionization starts from around 5 kV/cm, and the acceptors are completely ionized above 10 kV/cm. This value of the electric field is comparable to the static electric field necessary to ionize gallium acceptors, which is estimated as $I_p/ea_B^* \approx 10$ kV/cm (the binding energy $I_p$=11.3 meV and the effective Bohr radius $a_B^* \approx 10$ nm).[1,28] Thus, THz field induces a remarkable distortion of the Coulomb potential between the hole and the gallium ion, causing the bound hole to tunnel through the lowered potential barrier.

In order to treat the tunneling ionization with a strong THz electric field, it is necessary to solve the time-dependent Schrödinger equation. Actually the Keldysh parameter $K$,[29] an indicator to distinguish between the perturbative multi-photon ($K \gg 1$) and quasi-static tunneling regimes ($K \ll 1$), becomes unity when the applied electric field strength and frequency are respectively 4.6 kV/cm and 1 THz.[30] However, in the high field regime where ionization rate becomes large, we may apply the adiabatic approximation and assume that the released hole density can be estimated by using a rate equation with a static field-assisted tunneling time. Also, by neglecting the recombination process of free holes with ionized acceptors, since the 200-ps delay between pump and probe pulses is much shorter than the recombination time in Ge:Ga ~10 ns,[31,32] the time evolution of the bound hole density under THz pulse irradiation can be described by:

$$\frac{dN_b(t)}{dt} = -\,\tau_t^{-1}(t)N_b(t), \tag{2}$$

where $\tau_t(t)$ is the direct tunneling time from the acceptor state to the valence band. In the case of acceptors in semiconductors, the tunneling rate is given by:[33,34]

$$\tau_t^{-1}(t) = \omega_0 \left(\frac{6\alpha}{E_{THz}(t)}\right)^{2n_l^*-1} \exp\left(-\frac{\alpha}{E_{THz}(t)}\right), \tag{3}$$

where the following parameters defined in Ref. [34] are used: for the gallium acceptor in germanium, $\omega_0 = 6.7 \times 10^{13}\,\mathrm{s^{-1}}$, $n_l^* = 0.47$ and $\alpha = 16$ kV/cm. From Eq. (3), the tunneling time can be estimated to be ~0.5 ps at the electric field of 5 kV/cm.

By integrating Eq. (2), the remaining bound hole density can be represented as follows:

$$N_b^{cal} = N_b^0 - \int_0^\infty \tau_t^{-1}(t)N_b(t)\,dt. \tag{4}$$

Using the temporal profile of the pump electric field $E_{THz}(t)$ in Fig. 2(a), we can numerically calculate the bound hole density $N_b^{cal}$ from Eq. (4) without any fitting parameters. In Fig. 4, the theoretical curve obtained by using Eqs. (3) and (4) reproduces the experimental values over the whole range of electric field strengths. This result means that the direct field-assisted tunneling process dominates the ionization of the acceptor. As shown in the inset of Fig. 4, the field dependence of the bound hole density is not sensitive to the temperature difference. These results indicate that in the high field regime, the phonon-assist tunneling process negligibly contributes to the shallow impurity ionization.[5]

In summary, we have investigated the electric-field dependence of the THz optical constant of gallium-doped germanium by THz excitation and spectroscopy. With increasing peak electric field of the THz pump, the absorption peaks due to the internal acceptor levels disappear and the Drude dispersion appears in the THz frequency region. Good agreement between the experimental results and the theoretical calculations assuming a direct field-assisted tunneling process implies that the interaction between the impurity and the THz field enters the adiabatic direct tunneling regime.

We would like to thank Masaya Nagai for his help in the initial stage of this work and Susumu Komiyama for stimulating discussion. This study was supported by KAKENHI (Grant Nos. 23244065 and 20104007) from JSPS and MEXT of Japan. This work was partly supported by KAKENHI (Grant No. 24760042) from JSPS of Japan and the Grant-in-Aid for the Global COE Program "The Next Generation of Physics, Spun from Universality and Emergence" from MEXT of Japan.

**Figure Captions**

**Fig. 1** (a) Schematic configuration of THz excitation and spectroscopy. L: lens, BS: beam splitter, PM: off-axis parabolic mirror, WG: wire-grid polarizer, EOC: electro-optic crystal, BD: balanced detector, QWP: quarter wave plate, and WP: Wollaston prism. The configuration of the setup for THz generation is the same as in Ref. 15. The inset shows that the THz pulse is split into two pulses using partial reflection between WG1 and WG2. The pulse intensity is tuned by the rotation of WG3.

**Fig. 2** (a) THz field temporal profile and (b) power spectrum. (c) Real part of the conductivity of Ge:Ga at 15 K without the THz-pump excitation. The inset shows the primary excitation paths between the acceptor levels.

**Fig. 3** Real part of the dielectric constant and conductivity of Ge:Ga at 15 K for different THz electric fields: (a), (b) without pump excitation, (c), (d) $E_{THz}$=3.6 kV/cm, (e), (f) $E_{THz}$=13.5 kV/cm. Dashed lines show fitting curves with the Drude-Lorentz model described by Eq. (1).

**Fig. 4** Bound hole density after the THz-pump pulse irradiation. The symbols show experimental values deduced by fitting the optical constant (Figs. 3). The solid line is the theoretical calculation assuming the direct field-assisted tunneling process described by Eqs. (2)-(4). The inset shows the data measured at T=10 K and 20 K.

# Fig.1

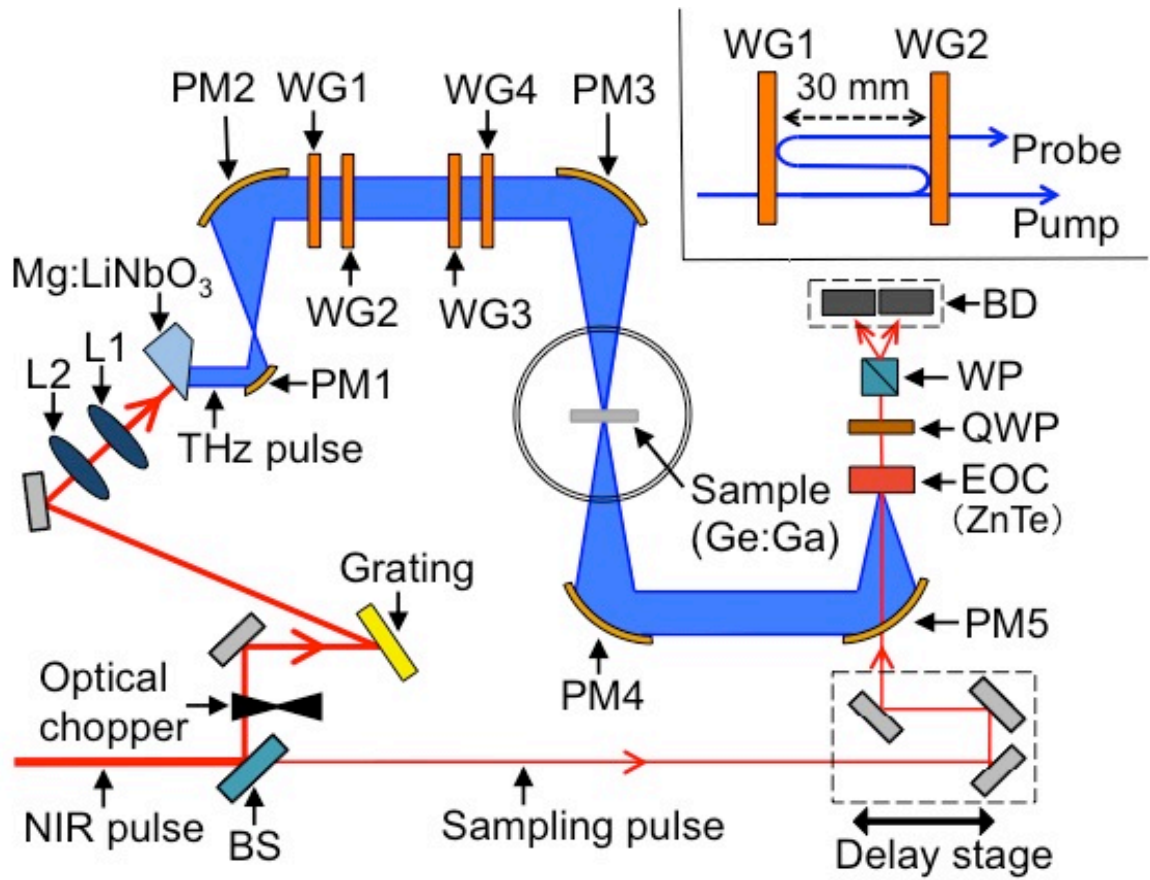

# Fig. 2

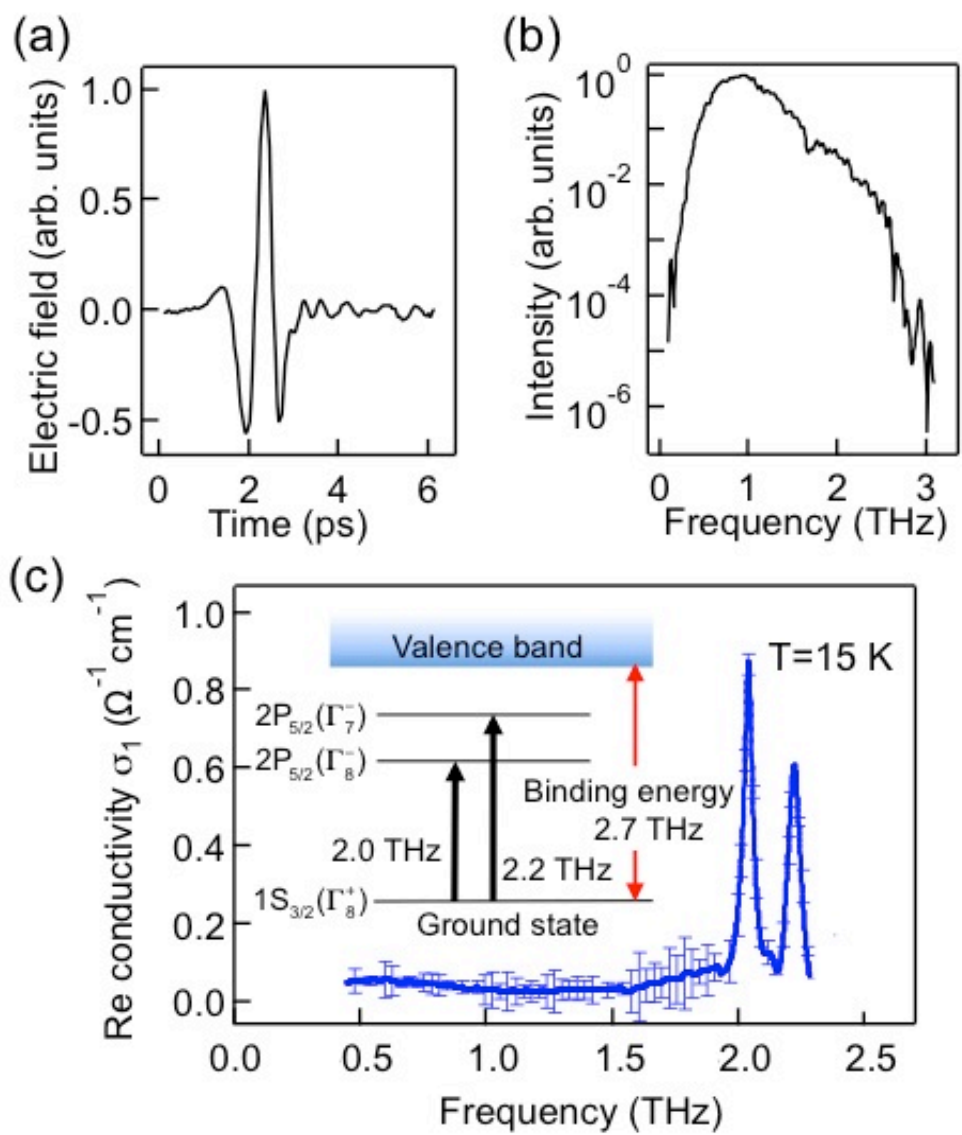

**Fig. 3**

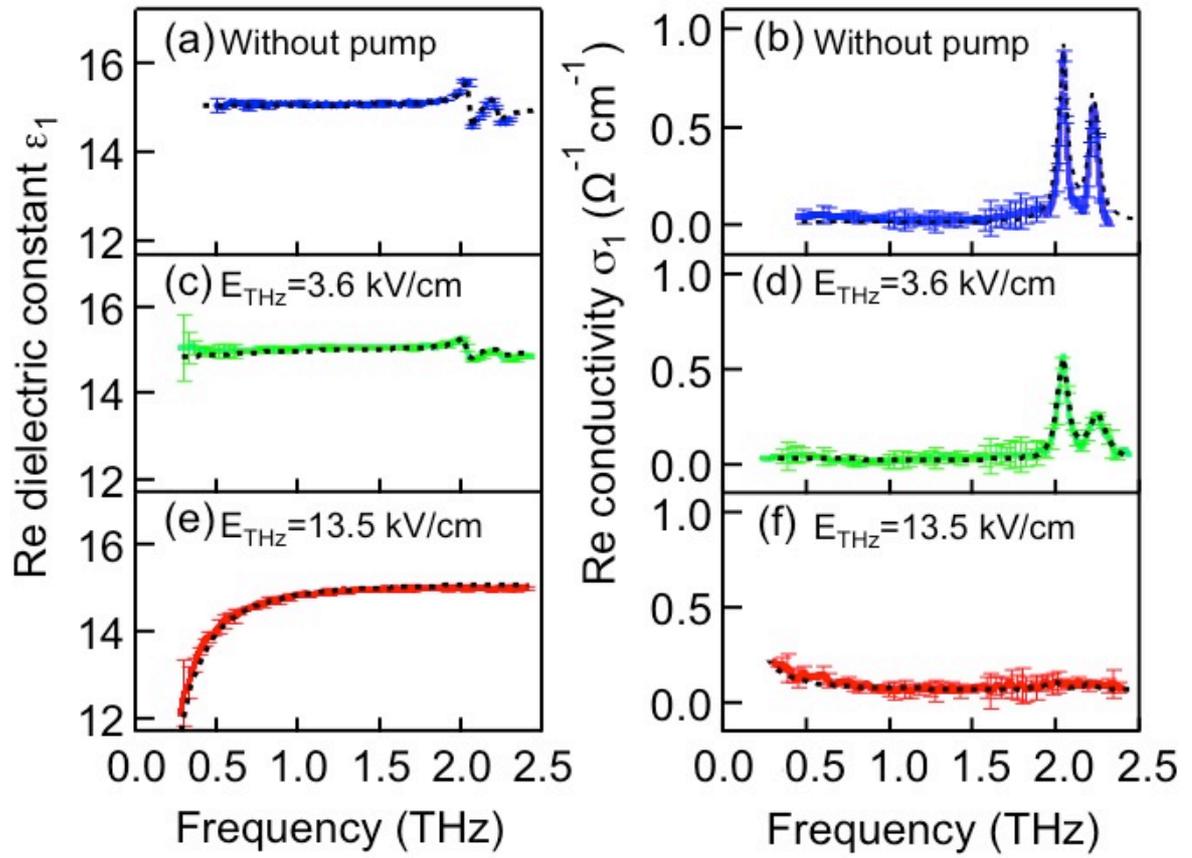

**Fig. 4**

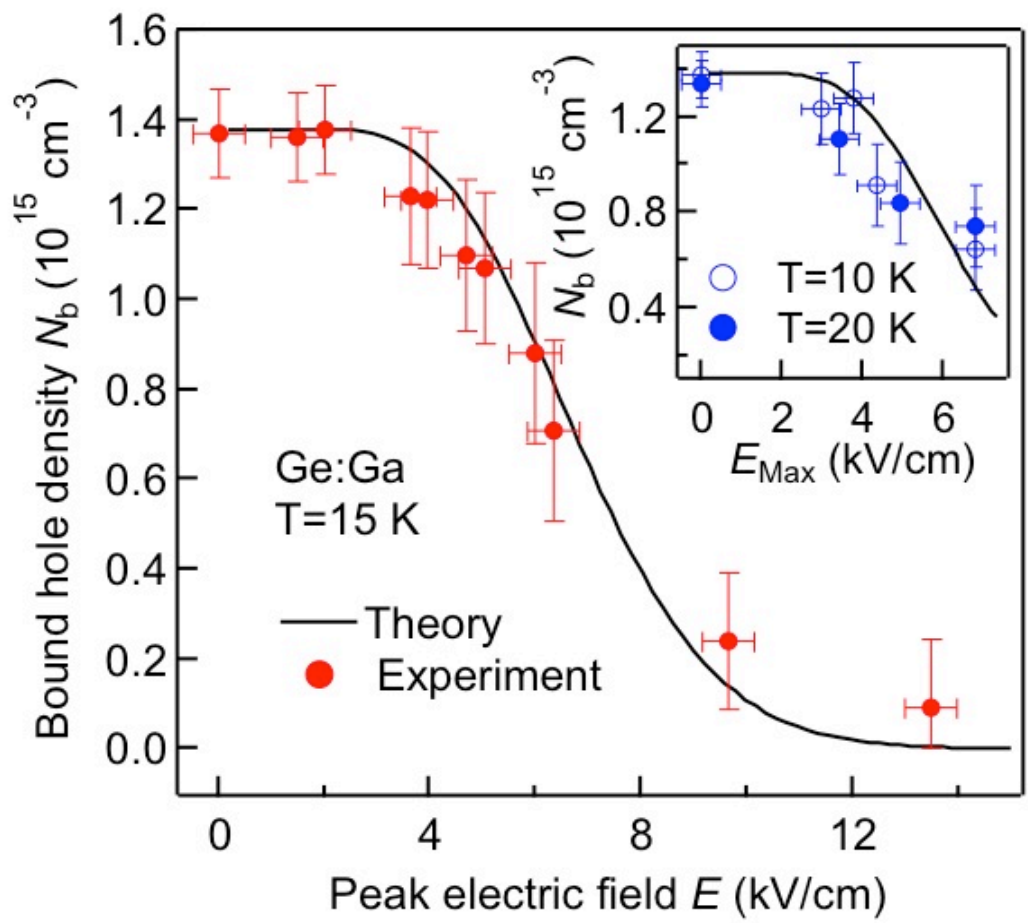